\documentclass{aa}
\usepackage{graphicx}
\usepackage{txfonts}
\usepackage{color}
\usepackage{natbib}
\bibpunct[; ]{(}{)}{,}{a}{}{,}
\def\asec{\ifmmode ^{\prime\prime}\else$^{\prime\prime}$\fi}

\def\it{\sl}
\def\degs{\ifmmode ^{\circ}\else$^{\circ}$\fi}
\def\amin{\ifmmode ^{\prime}\else$^{\prime}$\fi}
\def\asec{\ifmmode ^{\prime\prime}\else$^{\prime\prime}$\fi}
\def\fm{\hbox{$.\!\!^{\rm m}$}}            % Fractions of magnitudes
\def\fdg{\hbox{$.\!\!^\circ$}}          % Fractions of degrees
\def\farcs{\hbox{$.\!\!^{\prime\prime}$}}  % Fractions of arcseconds
\def\psr{PSR~J0357$+$3205}
\def\degs{\ifmmode ^{\circ}\else$^{\circ}$\fi}
\def\amin{\ifmmode ^{\prime}\else$^{\prime}$\fi}
\def\farcm{\hbox{$.\mkern-4mu^\prime$}}
\def\eqalign#1{\null\,\vcenter{\openup1\jot \m@th
   \ialign{\strut\hfil$\displaystyle{##}$&$\displaystyle{{}##}$\hfil
   \crcr#1\crcr}}\,}
\begin{document}

   \title{Deep optical observations of the $\gamma$-ray pulsar J0357$+$3205
\thanks{Based on observations made with the Gran Telescopio Canarias (GTC),
instaled in the Spanish Observatorio del Roque de los Muchachos
of the Instituto de Astrofísica de Canarias, in the island of La Palma
under Programme GTC3-12BMEX.
}
}

 %  \subtitle{I. Overviewing the $\kappa$-mechanism}
\author{A.~Kirichenko\inst{1,2}
\and A.~Danilenko\inst{1}
\and Yu.~Shibanov\inst{1,2}
\and P.~Shternin\inst{1,2}
\and S.~Zharikov\inst{3}
\and D.~Zyuzin\inst{1}
}

\offprints{Aida Kirichenko, \\  \email{aida.astro@mail.ioffe.ru}}

\institute{
Ioffe Physical Technical Institute, Politekhnicheskaya 26,
St. Petersburg, 194021, Russia \\
aida.astro@mail.ioffe.ru, danila@astro.ioffe.ru, shib@astro.ioffe.ru, pshternin@gmail.com, da.zyuzin@gmail.com
\and
St. Petersburg State Polytechnical University, Politekhnicheskaya 29,
St. Petersburg, 195251, Russia
\and
Observatorio Astron\'{o}mico Nacional SPM, Instituto de Astronom\'{i}a, Universidad Nacional
Aut\'{o}nomia de Mexico, Ensenada, BC, Mexico \\
zhar@astrosen.unam.mx
}

%   \date{Received September 15, 1996; accepted March 16, 1997}

% \abstract{}{}{}{}{}
% 5 {} token are mandatory

  \abstract
   {A middle-aged radio-quiet pulsar J0357+3205 was discovered  
   in gamma-rays with \textit{Fermi} and later in X-rays with \textit{Chandra} and
   \textit{XMM-Newton} observatories. It produces an unusual thermally-emitting pulsar 
   wind nebula observed in X-rays.
   }
   {
   Deep optical observations were obtained to search for the pulsar optical counterpart 
   and its nebula using the Gran Telescopio Canarias (GTC).
   }
   {
   The direct imaging mode  in the Sloan $g'$ band was used. 
   Archival X-ray data were reanalysed and compared with the optical data.
   }
   {
   No pulsar optical counterpart was detected  down to $g'$~$\geqslant$
   28\fm1.  No  pulsar nebula was either
   identified in the optical.  We confirm early results that the X-ray spectrum of
   the pulsar consists of a nonthermal power-law component of the pulsar
   magnetospheric origin dominating at high energies and a soft thermal 
   component from the neutron star surface.  Using  magnetised partially ionised 
   hydrogen atmosphere models in X-ray spectral fits 
   we found that the thermal component 
   can come from entire surface of the cooling neutron star with a temperature of 36$^{+9}_{-6}$ eV,
   making it one of the coldest among cooling neutron stars known. 
   The surface temperature agrees with the standard neutron star cooling scenario.
   The optical upper limit does not put any additional constraints 
   on the thermal component, however it implies  a strong spectral break for the nonthermal component
   between the optical and X-rays as is observed in other middle-aged pulsars. 
   }
   {
   The thermal emission from the entire surface of the neutron star 
   likely dominates over the nonthermal emission in the UV range.
   Observations of the \psr~in this range are promising 
   to put more stringent constraints on its thermal properties.
   }


\authorrunning{A.~Kirichenko et al.}
\titlerunning{Middle-aged $\gamma$-ray \psr }
   \maketitle

\section{Introduction}
\label{sec1} 
Gamma-ray pulsars are considered as one of the main targets 
of the \textit{Fermi} mission. For the five years of activity 
The Large Area Telescope
(LAT) has discovered numerous amounts of such sources 
previously observed in the radio band. But, apart from the 
ability to detect many known radio 
pulsars in $\gamma$-rays, \textit{Fermi} LAT also affords 
the opportunity to discover pulsars independently 
in so-called blind searches 
\citep[cf.][]{Saz09}. 
These blind searches were quite successful leading to 
discovery of about three dozens of pulsars in 
$\gamma$-rays \citep[see e.g.,][]{Saz13, Pletsch}. 
Further multiwavelength investigations of these objects 
are crucial for unveiling the pulsar 
emission nature. Because \textit{Fermi} pulsars are 
typically nearby and energetic \citep{Saz13}, they, in particular, 
appear to be promising targets for studies in X-ray and optical domains. 

A middle-aged radio-quiet PSR J0357$+$3205 with period $P$~= 444~ms, 
magnetic field $B$~= 2.3~$\times$~10$^{12}$~G
and characteristic age $P$/2$\dot{P}$~= 5.4~$\times$ 10$^{5}$ yr
was discovered in one of the \textit{Fermi} LAT blind frequency searches
\citep{BlindFreqAbdo2009}. The distance to the pulsar of 
about 500 pc was estimated by \citet{XrayFirstDeLuca} 
based on the $\gamma$-ray ``pseudo-distance'' 
relation \citep[see e.g.,][]{sazparkinson2010ApJ}. 
First X-ray observations of the pulsar field with \textit{Chandra} 
had revealed a faint X-ray counterpart
of the object with an extended (9 arcmin) X-ray tail \citep{XrayFirstDeLuca}. 
Subsequent \textit{XMM-Newton} observations had shown clearly that
emission from the pulsar itself is generally nonthermal with a soft thermal 
component \citep{XraySecondDeLuca}. The pulsar field was also observed 
in the optical and near-infrared bands with 2.5--4 m class telescopes. 
No counterpart was found down to $V \ga $ 26\fm7 \citep{XrayFirstDeLuca}.  

To search for an optical counterpart of PSR J0357$+$3205 
and/or its tail at a higher 
sensitivity level, we performed deep optical  
observations  with the 10.4 m GTC. 
The details of  observations and data reduction are described
in Sect.~\ref{sec2}, our results together with reanalysis of 
the archival X-ray data are presented in Sect.~\ref{sec3} and are discussed
in Sect.~\ref{sec4}.

%%%%%%%%%%%%%%%%%%%%%%%%%%%%%%%%%%%%%%%%%%%%%%%%%%%% fig. direct images
\begin{figure*}[t]
\setlength{\unitlength}{1mm}
\begin{center}
\begin{picture}(145,65)(0,0)
\put (-20,-20) {\includegraphics[width=100.0mm, clip=]{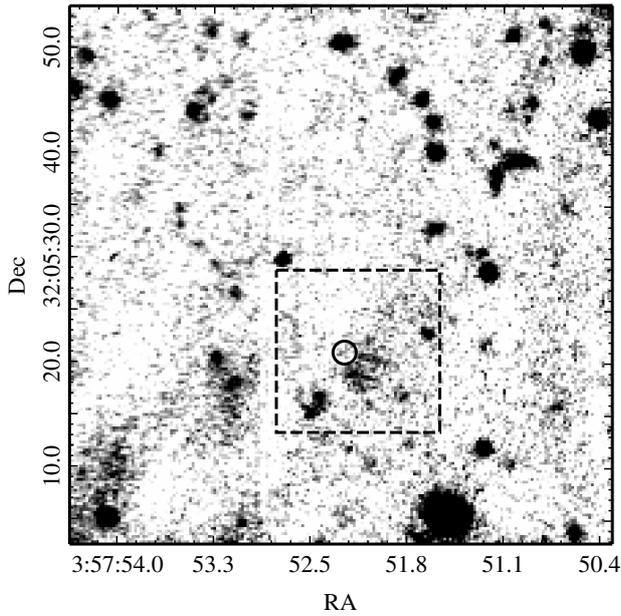}}
\put (80,-12.5) {\includegraphics[width=80mm, clip=]{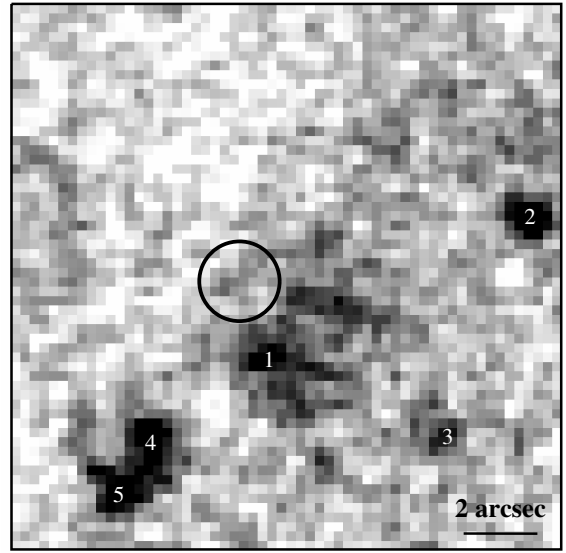}}
\end{picture}
\end{center}
\vspace{21mm}
\caption{{\sl Left} panel: GTC/OSIRIS $\sim$~51\asec~$\times$~51\asec~Sloan 
$g'$-image fragment of the \psr~field. 
The circle shows 3$\sigma$ X-ray pulsar position uncertainty
for the optical observations epoch (see text for details). The  
15\asec~$\times$~15\asec~pulsar vicinity within the dashed
rectangle is enlarged in the {\sl right} panel and smoothed 
with a one pixel Gaussian kernel.
The sources discussed in the text are labelled by numbers.   
 }
\label{fig:1}
\end{figure*}
%%%%%%%%%%%%%%%%%%%%%%%%%%%%%%%%%%%%%%%%%%%%%%%%%%%%%%%%%%%%%%%%%%%%%%%%%%%%%%

\section{GTC data}
\label{sec2}

%%%%%%%%%%%%%%%%%%%%%%%%%%%%%%end Table 1 %%%%%%%%%%%%%
\begin{table}[t]
\caption{Log of the GTC/OSIRIS observations of
\object{\psr}. }
\begin{center}
\begin{tabular}{lllll}
\hline\hline
   Date           &  Band         & Exposure               &  Airmass      & Seeing          \\
                  &               & [s]                    &               & [arcsec]        \\
\hline \hline
   2012-12-06     &  $g'$          & 700~$\times$~4           & 1.07          &  0.8      \\
   2012-12-16     &  $g'$          & 700~$\times$~4           & 1.13          &  0.9--1.0      \\
   2013-01-05     &  $g'$          & 700~$\times$~2           & 1.09          &  0.9--1.0      \\
   2013-01-13     &  $g'$          & 700~$\times$~4           & 1.18          &  0.8          \\
\hline
\end{tabular}
\end{center}
\label{t:log}
\end{table}
%%%%%%%%%%%%%%%%%%%%%%%%%%%%%%end Table 1 %%%%%%%%%%%%%

%%%%%%%%%%%%%%%%%%%%%%%%%%% OBSERVATIONS
\subsection{Observations and data reduction}
%%%%%%%%%%%%%%%%%%%%%%%%%%%%
\label{obs}
The observations of the pulsar field were carried out in the Sloan $g'$ band with the Optical System for
Imaging and low-intermediate Resolution Integrated Spectroscopy
(OSIRIS\footnote{For instrument features see {http://www.gtc.iac.es/instruments/osiris/}.})
at the GTC in a queue-scheduled service mode in 2012 December and 2013 January. With the image scale
of 0\farcs254/pixel (2$\times$2 binning) and unvignetted field size
of 7\farcm8~$\times$~7\farcm8 available with the OSIRIS detector consisting of  a mosaic of two CCDs,
we obtained four sets of 700-second dithered exposures in a grey time.
The pulsar was exposed on  CCD1.
The observing conditions were 
photometric,
with seeing  varying
from 0\farcs8 to 1\farcs0 (see Table~\ref{t:log}).

Standard data reduction, including bias subtraction, flat-fielding, 
cosmic-ray removal, and bad pixel correction,
was performed with {\tt IRAF} and {\tt MIDAS} tools.
Subsequent data inspection showed that each single exposure was 
significantly contaminated by a nonuniform background.
This was caused by the fact that OSIRIS detector filter wheels are 
inclined at an angle of 10\fdg5 with respect to
the incident light. For this reason, the central wavelength of the 
filter moves slightly bluewards from CCD1 to CCD2
of the OSIRIS focal plane. For the broad-band filters this effect 
is small\footnote{It leads to a displacement of as much as 30 \AA\ for the
Sloan $g'$ filter, see OSIRIS user manual for details: 
{http://www.gtc.iac.es/ instruments/osiris/media/ OSIRIS-USER-MANUAL\_v2.1.pdf}.}, 
but becomes visible in case of a high background as on 
our images obtained during grey time.
In order to eliminate the contamination we performed 
illumination correction 
for each observational set.

Finally, using a set of unsaturated stars, we aligned all the corrected individual exposures
to the best one obtained in the highest quality seeing conditions. 
The alignment accuracy was $\la$~0.1 pixel.
As a result, we obtained a combined image with mean seeing of 0\farcs9, 
mean airmass of 1.12 and total integration time of 9.8 ks.

\subsection{Astrometric referencing and photometric calibration}
%%%%%%%%%%%%%%%%%%%%%%%%%%%%%%%%%%%%%%%%%%%%%
\label{astroref}

In order to perform a precise astrometric referencing we used the positions 
of the astrometric standards from the USNO-B1
astrometric catalogue\footnote{See {http://www.nofs.navy.mil/data/fchpix/}.}. 
A set of 10 isolated unsaturated stars was selected on
the combined image. Precise pixel coordinates of these stars were obtained 
using the {\tt IRAF} task {\sl imcenter} with
an accuracy of $\la$~0.003 pixel. For the astrometric transformation 
we applied the {\tt IRAF} task {\sl ccmap}.
Formal {\sl rms} uncertainties of the astrometric fit were $\Delta$RA~$\la$ 0\farcs123
and $\Delta$Dec~$\la$ 0\farcs155, which is consistent with the nominal catalogue uncertainty
of $\approx$~0\farcs2. The resulting conservative 1$\sigma$ 
referencing uncertainty for the combined images is
$\la$~0\farcs23 for RA and $\la$~0\farcs25 for Dec.

For photometric calibration we used the G158-100 Sloan standard
\citep{Smith} observed the same nights as our target. 
The atmospheric extinction  coefficient for $g'$ taken from the OSIRIS user manual 
is 0.16(1) mag~airmass$^{-1}$.
The determined magnitude
zero-point for our $g'$ image is 28.64(5).

%%%%%%%%%%%%%%%%%%%%%%%%%%%%%%%%%%%%%%%%%%%%%%%%
\section{Results}
%%%%%%%%%%%%%%%%%%%%%%%%%%%%%%%%%%%%%%%%%55
\label{sec3}
\begin{table*} %%%[b]
  \caption{Best-fit  parameters of the pulsar X-ray spectrum with three models. 
  Temperatures $T^{\infty}$ and emitting area radii $R^{\infty}$ 
  are  as measured by a distant observer.
  $N_H$ is the absorbing column density. PL$_{norm}$ and $\Gamma$ 
  are PL normalisation and photon spectral index. 
  Errors are at 90\% confidence.}
  %\begin{center}
  \begin{center}
  \begin{tabular}{ccccccc}
  %\hline
  \hline
  %\multicolumn{8}{c}{ }  \\
  Model    & $N_H$               & $\Gamma$            & PL$_{norm}$                                &   $T^{\infty}$       &  $R^{\infty}$       & $\chi^2$/dof (dof) \\ % &  $T_s$   \\    
           & 10$^{21}$ cm$^{-2}$ &                     & 10$^{-5}$ ph keV$^{-1}$~cm$^{-2}$~s$^{-1}$ &   eV                 &  d$_{500pc}$ km     &                     \\ %  &  eV   &                     \\                
  \hline
  \multicolumn{7}{c}{ }  \\
  BB+PL    & $1.4^{+0.5}_{-0.4}$ & $2.2^{+0.2}_{-0.2}$ & $1.1^{+0.2}_{-0.2}$                        & $93^{+9}_{-9}$       & 0.5$^{+0.4}_{-0.2}$ & 1.05 (244) \\      %& \\ % $\la$~31 & 
  \multicolumn{7}{c}{ }  \\
  NSA+PL   & $1.6^{+0.9}_{-0.6}$ & $2.1^{+0.3}_{-0.3}$ & $1.0^{+0.3}_{-0.3}$                        & 37$^{+12}_{-8}$ & 6$^{+10}_{-5}$ & 1.05 (244) \\        %$\la$~20
  \multicolumn{7}{c}{ }  \\
  NSA+PL   & $2.4^{+0.2}_{-0.3}$ & $2.3^{+0.1}_{-0.2}$ & $1.2^{+0.2}_{-0.1}$                        & 30$^{+1}_{-1}$ & 15.73          & 1.07 (245) \\        %$\la$~20
  (fixed NSA normalisation)  & \multicolumn{6}{c}{ }  \\
  \multicolumn{7}{c}{ }  \\
  NSMAX+PL & $1.7^{+0.6}_{-0.5}$ & $2.0^{+0.2}_{-0.2}$ & $0.9^{+0.2}_{-0.2}$                        & 36$^{+8}_{-6}$ & 8$^{+12}_{-5}$         & 1.05 (244) \\        %$\la$~20
  \multicolumn{7}{c}{ }  \\
  NSMAX+PL & $2.1^{+0.2}_{-0.2}$ & $2.0^{+0.2}_{-0.2}$ & $1.0^{+0.1}_{-0.2}$                        & 31$^{+1}_{-1}$ & 15.73         & 1.05 (245) \\        %$\la$~20
  (fixed NSMAX normalisation)  & \multicolumn{6}{c}{ }  \\
  \hline
\end{tabular}
\end{center}
\label{t:x-fit}
\end{table*}

\subsection{The pulsar field}
\label{pulsar_vic}
In the {\sl left} panel of Fig.~\ref{fig:1} we present the resulting GTC $g'$ 
image fragment that contains PSR J0357$+$3205.
The circle is centred at the expected pulsar position 
with RA~= 03:57:52.293 and Dec~= +32:05:20.970 ($l$~= 162.76$^{\circ}$ and $b$~= $-$16.01$^{\circ}$)
for the GTC observations epoch. 
It was estimated using the X-ray pulsar position obtained in the \textit{Chandra} 
2009 observations \citep[see][]{XrayFirstDeLuca} 
and accounting for the pulsar proper motion of 0\farcs165 $\pm$ 0\farcs030 yr$^{-1}$
\citep{Fast} at $\sim$~3.2 yr time base between the \textit{Chandra} 2009 and GTC 2012--2013 observations. 
The circle radius of $\sim$~1\farcs1 corresponds to
the 3$\sigma$  pulsar position uncertainty in the optical image,
that accounts for the optical astrometric referencing, proper motion,
and pulsar X-ray position uncertainties\footnote{We used the \textit{Chandra} 
pulsar positional error of 0\farcs25 from \citet{XrayFirstDeLuca}.}.

The   pulsar vicinity marked by the dashed rectangle is enlarged
in the {\sl right} panel of Fig.~\ref{fig:1}. 
In this region using the {\tt IRAF} task {\sl daofind}  we find five 
compact sources detected at $\ga$3$\sigma$ significance, 
they are labelled by numbers. 
Source ``1'' has a $g'$ magnitude of 26\fm6(1) and 
is the closest object to the pulsar. However, it locates 2\farcs2 away from the X-ray  position, 
implying the offset significance of $\sim$~6$\sigma$. Such a high displacement rules 
out this object  
as a pulsar optical counterpart.
Based on its spatial brightness profile we cannot firmly distinguish 
if it is a point source or a bright part of an extended 
structure immediately SW of the pulsar. 
Objects ``2'' and ``3'' with magnitudes of 26\fm4(1) and 26\fm8(2), 
respectively, are likely point-like sources. 
They are even more distant from the pulsar position and are certainly unrelated objects. 
Finally, extended sources ``4'' and ``5'' with magnitudes 25\fm9(1) 
and 26\fm3(1) are probably galaxies or 
irrelevant blended stellar objects.
A compact flux enhancement is seen within the 3$\sigma$ pulsar position error circle. 
However, it does not exceed significantly 
the background fluctuations in this area, so currently we do not have 
any strong arguments to propose this as a real object.

Therefore, based on our optical  data, we can give  only a 
conservative estimation on the optical flux upper limit 
from the pulsar.
Following a standard procedure \citep[e.g.,][]{zhar2} we obtained the point 
source 3$\sigma$ flux upper limit of $\la$ 0.023 $\mu$Jy ($g'\ga$ 28\fm1).
This is currently the most stringent constraint on the optical flux of the pulsar.

We did not detect any reliable ($\ga 3\sigma$) extended emission except the feature 
around the source ``1'' immediately SW of the pulsar.
This feature cannot be associated with the long SE tail behind the pulsar detected in X-rays.
However, 
the brightest parts of the tail in X-rays
are outside the GTC field of view. 
The nature of the extended SW feature 
is unclear. 
It could be  an interstellar cloud 
in the pulsar vicinity partially ionised
by the pulsar emission and emitting in [OIII]5007/4969A lines
which fall within the $g'$ bandpass. Observations in other broad
and/or narrow bands would be useful to understand the origin of
the source.

\subsection{Reanalysis of the X-ray spectrum}
\label{s:x-ray}
To evaluate how informative the pulsar flux upper limit is, it 
is useful to compare it with X-ray spectral data. To do that, we performed an
independent X-ray data analysis.
We retrieved  all the available archival X-ray data obtained with
\textit{Chandra}\footnote{ACIS-S, Obs. IDs 11239, 12008 and 14207 
(dates 2009.10.26, 2009.10.25 and 2011.12.24),
Exp. time 46 ks + 29 ks + 29 ks, PI A. de Luca.} 
and \textit{XMM-Newton}\footnote{EPIC-MOS and PN, Obs. ID 0674440101,
date 2011.09.15, exp. time 110 ks, PI A. de Luca.}.
To extract the pulsar spectra, we used 2\asec~and 30\asec~apertures
centred at the pulsar position,  which enclose $\ga$~90\% of the pulsar emission 
in the \textit{Chandra} and
\textit{XMM-Newton} data, respectively. The \texttt{CIAO} \texttt{v.4.5} {\sl specextract } 
and  \texttt{SAS} \texttt{13.0} {\sl especget } tools were used
for the extraction resulting in $\sim$3500 \textit{XMM-Newton}/EPIC 
and $\sim$1000 \textit{Chandra}/ACIS source counts.
Using the \texttt{XSPEC} \texttt{v.12.8.1} we then 
fitted the spectra in 0.3--10 keV range by an absorbed spectral model containing 
power-law (PL) and thermal emission components 
originating from the magnetosphere and the surface of the neutron star (NS), 
respectively.  We used the \texttt{XSPEC} photoelectric absorption model \texttt{phabs} 
with default abundances \texttt{angr} \citep{anders1989GeCo} and cross-sections \texttt{bcmc} 
\citep{bcmc1992ApJ}.
We also tried other abundances and cross-sections available 
in \texttt{XSPEC}\footnote{see http://heasarc.nasa.gov/xanadu/xspec/manual/XSmodelPhabs.html}, 
but this did not significantly change 
fit statistics ($\chi^{2}$) and fit parameters 
remained within their confidence intervals.
For the thermal  component we  
used either blackbody (BB) or magnetised neutron star hydrogen atmosphere 
NSA and NSMAX models \citep{pavlov1995lns,ho2008ApJS}, 
which provided equally acceptable fits. 

The best-fit parameters for the absorbed BB+PL, NSA+PL and NSMAX+PL models
are presented in Table~\ref{t:x-fit} where errors are at the 90\% confidence.
For the NSA model
we fixed NS mass $M$,  circumferential radius $R$, and 
surface magnetic field $B$ at 1.4M$_{\odot}$,  
13 km, and 10$^{12}$ G, respectively.  For the NSMAX model 
we fixed the redshift parameter $1+z$ at the value of 
1.21 which corresponds to the same $M$ and $R$. 
We selected the model 1200 from NSMAX family, which represents
the atmosphere with $B$~=10$^{12}$~G. Due to the 
space-time curvature near the NS, its apparent 
radius is $R(1+z)=15.73$~km. 
The values of temperatures $T^{\infty}$ and 
thermally emitting area radii $R^{\infty}$ 
(in units of d$_{500pc}$ km) are given as measured by a distant observer. 
For all models $R^{\infty}$ is derived from the model 
normalisation. 

\begin{figure}[t]
  \begin{center}
  \setlength{\unitlength}{1mm}
  \resizebox{15.5cm}{!}{
  \begin{picture}(50,24.5)(0,0)
    \put (-2,0)
    {\includegraphics[scale=0.162]{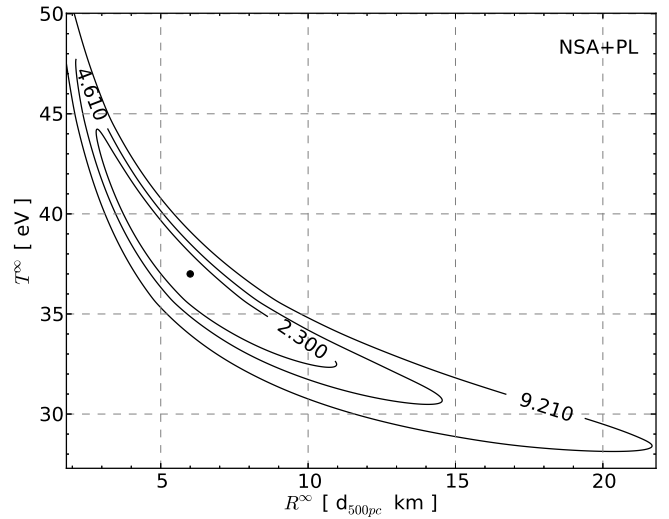}}
  \end{picture}
  }
  \end{center}
  \caption{
  68\%, 90\%, and 99\% 
  ($\Delta\chi^{2}=\chi^{2}-\chi^{2}_{min}$ = 2.3, 4.61, and 9.21, respectively) 
  confidence contours  
  of the neutron star effective  temperature vs.
  emitting area radius for the absorbed NSA+PL model.
  The dot shows the best-fit parameters (see Table~\ref{t:x-fit}).
  }
  \label{fig:contours}
\end{figure}

\begin{figure}[t]
  \begin{center}
  \setlength{\unitlength}{1mm}
  \resizebox{15.5cm}{!}{
  \begin{picture}(50,24.5)(0,0)
    \put (-2,0)
    {\includegraphics[scale=0.162]{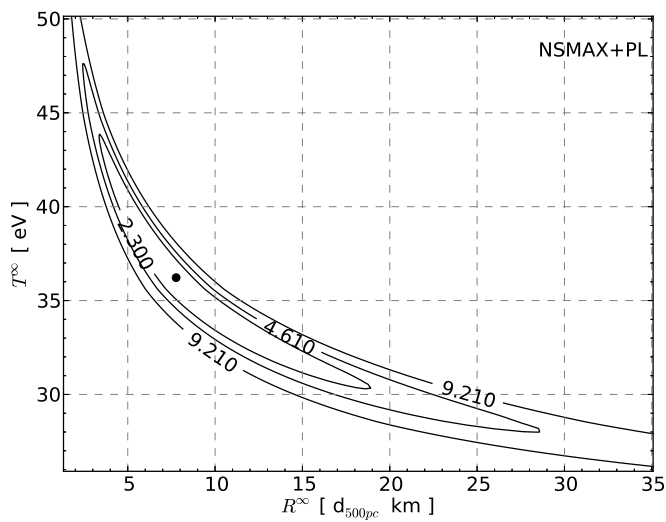}}
  \end{picture}
  }
  \end{center}
  \caption{
  The same as in Fig.~\ref{fig:contours} but for the absorbed NSMAX+PL model.  
  }
  \label{fig:contours_nsmax}
\end{figure}

The BB emitting area
is significantly smaller than the entire surface of the NS, 
but is consistent with a canonical pulsar polar 
cap size \citep{sturrock1971ApJ} 
of about 0.32 km derived  for the 13 km NS with the period of 444 ms.
At the same time, both NSA and NSMAX models 
give emitting area radii, which are much larger than the   
cap size, but agree well with the 
standard apparent NS size (10--20~km), especially accounting for 
the distance uncertainties (see below). 
This is also demonstrated by the $T^{\infty}-R^{\infty}$ confidence contours 
presented in
Figs.~\ref{fig:contours} and \ref{fig:contours_nsmax}. 
Moreover, if the normalisations for both hydrogen atmosphere models
are fixed according to apparent 
emitting area radius of the 15.73 km at 500 pc distance, 
the fit is still statistically acceptable 
(third and fifth rows in Table~\ref{t:x-fit}). 
While both NSA and NSMAX models 
give comparable best-fit parameters (see Table~\ref{t:x-fit}),
the NSMAX results appear to be more plausible, since this model
accounts for the partial ionisation of atmosphere plasma.
The latter is essential for magnetised hydrogen 
atmospheres with effective temperatures less than 90 eV 
\citep{potekhin1999PhRvE}.

\begin{figure}[t]
  \begin{center}
  \setlength{\unitlength}{1mm}
  \resizebox{15.5cm}{!}{
  \begin{picture}(50,25.5)(0,0)
    \put (-2,0)
    {\includegraphics[scale=0.165]{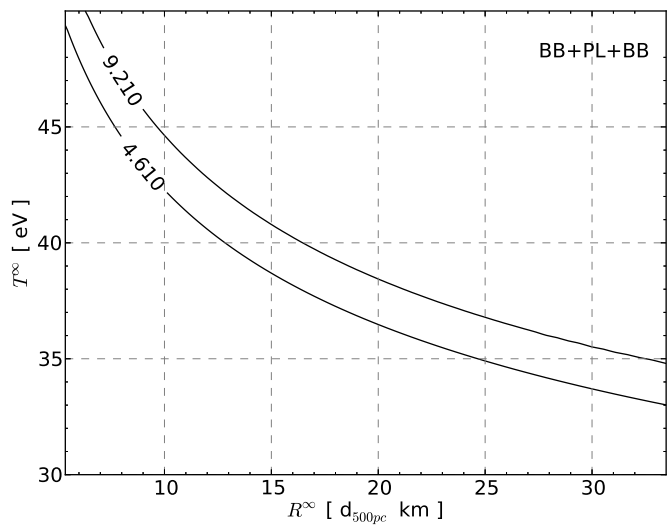}}
  \end{picture}
  }
  \end{center}
  \caption{
  90\% and 99\% ($\Delta\chi^{2}=\chi^{2}-\chi^{2}_{min}$ = 4.61 and 9.21, respectively) 
  upper confidence boundaries of the effective temperature vs.
  emitting area radius of the second BB component added to the absorbed BB+PL model.  
  }
  \label{fig:weiss}
\end{figure}

We also estimated an upper limit on 
the star entire surface temperature for the BB+PL model 
following the procedure 
used by \citet{weiss2011ApJ} for the Crab pulsar. 
We added an additional BB component to this model which 
is not required to describe the data and does not affect 
the initial best-fit. The upper limit 
is then derived from the upper boundary of the 
$R^{\infty}$--$T^{\infty}$ confidence contours for the new component.
Respective contours for 90\% and 99\%  confidence   
are shown in Fig.~\ref{fig:weiss}. 
Here the $R^{\infty}$ scale corresponds to a
reasonable range of the NS radii accounting for uncertainties of
the distance to the pulsar (see below).
For the $R^{\infty}$~= 15.73~$d_{500pc}$~km, 
the entire NS surface temperature
upper limit is 40 eV (at 99\% confidence). 

Our results for the BB+PL model are similar to those by \citet{XraySecondDeLuca}. 
However, the results for the NSA+PL model and for the upper limit on 
the surface temperature in the blackbody model
are different. The reason for the former is in the different value 
of magnetic field used. We checked that for $B=10^{13}$~G our 
results agree with \citet{XraySecondDeLuca}. 
In our spectral fits we use $B=10^{12}$~G, which is closer to the value 
inferred from the spindown measurements.
The blackbody upper limit reported by \citet{XraySecondDeLuca} 
is 38~eV for the NS radius 
of 10~km and distance of 500~pc. 
According to Fig.~\ref{fig:weiss} the upper limit for these parameters 
($R^\infty\approx13$~km for 1.4$M_\odot$ NS) should be 42 eV. 
Unfortunately, \citet{XraySecondDeLuca} do not
describe the method for obtaining their value and the reason 
for this discrepancy is unclear.

It is important to stress, that 
the hydrogen atmospheric and blackbody models have equal rights to be considered 
as the  interpretation of the thermal component 
of the emission. The blackbody model can mimic the emission 
from the iron atmosphere or from the condensed 
surface of the NS \citep[see e.g.,][and references therein]{vanadelsberg2005ApJ}. 
Deeper X-ray observations allowing for a phase resolved 
spectral analysis will enable us to distinguish between the models.
Note, that the parameters of the PL component, dominating at energies ~$\ga$ 1 keV,
are almost independent of the type of the thermal component  involved into 
the combined model (Table~\ref{t:x-fit}).

Finally, any reasonable single spectral model is not acceptable.
For instance, 
absorbed PL, BB, and NSA models 
give reduced $\chi^{2}$/(dof) of 1.30/(246), 2.52/(246), and 2.22/(246), respectively.
Any combined model with the PL replaced by a second  thermal component, 
which may represent the emission 
from two areas of the NS with different temperatures, 
is not acceptable either, e.g., an absorbed BB+BB, NSA+NSA, and  NSMAX+NSMAX give  
$\chi^{2}$/(dof)~= 1.27/(244), 1.68/(244), and 2.50/(244), respectively.

%%%%%%%%%%%%%%%%%%%%%%%%%%%%%%%%%%%%%%%%%%%%%% Fig. 8  %%%%%%%%%%%%%%%%%%%%%%%%%%
%%%%%%%%%%%%%%%%%%%%%%%%%%%%%%%%%%%%%%%%%%%%%%%%%%%%%%%%%%%%%%%%%%%%%%%%%%%%%%%%%%

%%%%%%%%%%%%%%%%%%%%%%%%%%%%%%%%%%%%%%%%%%%%%% Fig. 8  %%%%%%%%%%%%%%%%%%%%%%%%%%
%%%%%%%%%%%%%%%%%%%%%%%%%%%%%%%%%%%%%%%%%%%%%%%%%%%%%%%%%%%%%%%%%%%%%%%%%%%%%%%%%%

%%%%%%%%%%%%%%%%%%%%%%%%%%%%%%%%%%%%%%%%%%%%%% Fig. 8  %%%%%%%%%%%%%%%%%%%%%%%%%%
%%%%%%%%%%%%%%%%%%%%%%%%%%%%%%%%%%%%%%%%%%%%%%%%%%%%%%%%%%%%%%%%%%%%%%%%%%%%%%%%%%

%%%%%%%%%%%%%%%%%%%%%%%%%%%%%%%%%%%%%%%%%%%%%%%%%
\section{Discussion}
%%%%%%%%%%%%%%%%%%%%%%%%%%%%%%%%%%%%%%%%%%%%%%%%%%5
%%%%%%%%%%%%%%%%%%%%%%%%%%%%%%%%%%%%%%%%%%%%%% Fig. 8  %%%%%%%%%%%%%%%%%%%%%%%%%%
\begin{figure}[t]
  \begin{center}
  \setlength{\unitlength}{1mm}
  \resizebox{15.5cm}{!}{
  \begin{picture}(50,24.5)(0,0)
    \put (-1,0)
    {\includegraphics[scale=0.127]{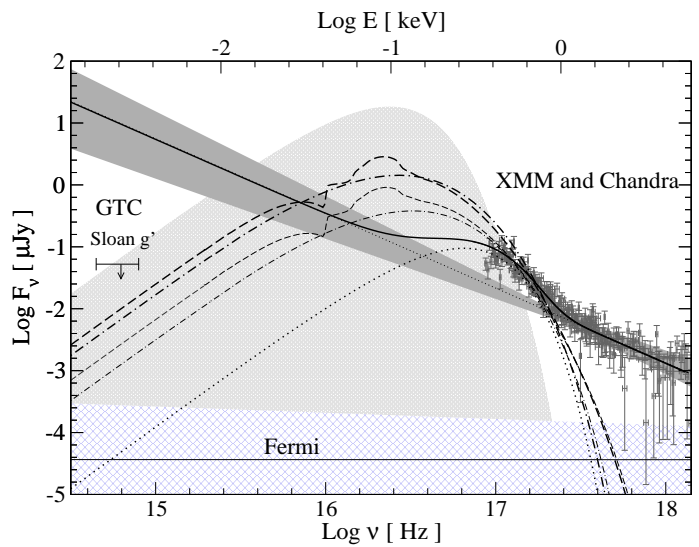}}
  \end{picture}
  }
  \end{center}
  \caption{
  Unabsorbed spectrum of PSR J0357$+$3205 in the optical-X-ray range. 
  The GTC $g'$-band flux upper limit 
  is indicated by the bar with an arrow.
  The \textit{XMM-Newton} and \textit{Chandra} data points  
  are shown by grey bars.\protect\footnotemark\
  The thick solid line is the best-fit BB+PL model to the X-ray data. 
  The thin dotted line with the dark-grey filled region is the PL component with 90\% uncertainties. 
  The thick dotted line is the BB hot spot 
  component. The thin/thick dash-dotted and dashed lines are 
  the best-fit NSA and NSMAX components  
  with variable/fixed normalisations. 
  The upper limit on the thermal spectral flux from the entire surface of the NS 
  obtained with the BB+PL+BB model 
  is shown by the light-grey filled region.
  The extrapolation of the $\gamma$-ray \textit{Fermi} spectrum with its uncertainties 
  is shown by the thin solid line with a hatched region.
  }
  \label{fig:spectrum}
\end{figure}
%%%%%%%%%%%%%%%%%%%%%%%%%%%%%%%%%%%%%%%%%%%%%%%%%%%%%%%%%%%%%%%%%%%%%%%%%%%%%%%%%%

From three statistically acceptable X-ray models, BB+PL, NSA+PL, 
and NSMAX+PL, considered in the previous Section, 
two latter result in similar parameters of the pulsar thermal emission. 
However, the NSMAX model is more justified from the 
physical reasons. Therefore in what follows we omit the NSA+PL 
model for simplicity.

In order to compare the X-ray data and the optical upper limit, 
the latter must be corrected for the interstellar extinction 
$A_V$. The standard $A_V$--$N_H$ relation \citep{predehl1995AsAp} can be used 
to estimate $A_V$.
The BB+PL and NSMAX+PL X-ray spectral fits suggest  
$N_H$ in a range of (1.0--2.3)~$\times$ 10$^{21}$ cm$^{-2}$ (Table~\ref{t:x-fit}). 
This corresponds 
to the $A_V$ range of 0.6$-$1.4. However, $A_V$ can hardly 
exceed        
the entire Galactic extinction in this direction of 0.8 recently 
estimated by \citet{Schlaf}. Therefore,  
we accept 0.8 as a conservative extinction value for dereddening the optical upper limit.
At the same time, the actual $N_H$ value can be larger than 
1.4~$\times$ 10$^{21}$ cm$^{-2}$ 
which corresponds 
to this $A_V$. For instance, 
\citet{XraySecondDeLuca} estimate entire Galactic
$N_H=(2.1\pm0.2)\times 10^{21}$~cm$^{-2}$ from the spectral 
analysis of extra-galactic sources 
in the pulsar field.

Considering the half-thickness of $\sim$~100 pc for the Galactic gaseous disk responsible 
for the extinction, the pulsar latitude $b$~= $-$16$^{\circ}$, and the minimal 
$A_V=0.6$ ($N_H=10^{21}$~cm$^{-2}$) we obtain 
the minimum distance to the pulsar of $\sim$~270 pc. This value is derived assuming the 
uniform $A_V$ scaling with distance within the disk.
The upper limit on
the distance of $\sim$900 pc was estimated by \citet{XraySecondDeLuca}
based on an assumption 
that the pulsar intrinsic $\gamma$-ray
luminosity cannot exceed its spin-down energy loss rate.       
The 500 pc distance accepted in Sect. 3.2 is consistent with these limits.

At the distance of 500 pc the optical (in the $V$ band) and 
nonthermal X-ray (in the range of 2--10 keV) 
luminosities of the
pulsar are $L_{V}$~$\la$ 1.1~$\times$~10$^{27}$ erg~s$^{-1}$ and
$L_{X}$~= 6.0~$\times$~10$^{29}$ erg~s$^{-1}$. Accounting
for the spin-down luminosity  $\dot E= 5.8 \times 10^{33}$
erg~s$^{-1}$ \citep{BlindFreqAbdo2009} they yield  the optical
and X-ray efficiencies of the pulsar  
$L_{V}/\dot E$~$\la$
10$^{-6.7}$ and $L_{X}/\dot E$~$\approx$ 10$^{-4.0}$.
These values
are compatible with the empirical X-ray luminosity and efficiency
{\sl vs} age dependencies demonstrated by the pulsars detected in
the optical and X-rays \citep[e.g.,][]{zharikov2006AdSpR, zhar2}. This also 
supports the distance estimate of 500 pc.

In Fig.~\ref{fig:spectrum} we compare the unabsorbed X-ray 
and $\gamma$-ray spectra of the pulsar with 
the optical upper limit of 0.052 $\mu$Jy derived from the GTC observations and dereddened 
with $A_V$~= 0.8.
\footnotetext{The data were unfolded and 
unabsorbed   
by applying the factor (unfolded unabsorbed model)/(folded absorbed model) 
in each spectral bin, assuming the best-fit BB+PL model. 
This procedure is analogous to how an unfolded spectrum is plotted by
the \texttt{Xspec} \texttt{plot ufspec} command.
Note that the data points obtained this way are model-dependent and for different models will follow 
the respective best-fit lines. 
}
The optical upper limit is 
two orders of magnitude lower than 
it would be expected from 
the extrapolation of the PL spectral component to the optical range.
According to Sect.~\ref{s:x-ray} the PL component is essential 
to describe the high-energy tail
of the pulsar X-ray spectrum.
This suggests a spectral break in the PL component  between the optical and X-rays, 
as it is observed for all middle-aged pulsars detected in both domains 
\citep{shibanov2006AsAp}. The extrapolation of the $\gamma$-ray PL spectrum 
of the pulsar (see Fig.~\ref{fig:spectrum})
lies well below the optical upper limit. 

The best-fit NSMAX spectral components with variable and 
fixed normalisations are shown by thin and thick dashed lines in Fig.~\ref{fig:spectrum}, 
respectively. For completeness we 
also show the NSA spectral components (dash-dotted lines). 
As seen, to significantly constrain the NS thermal emission in these models
one has to go as deep as $\sim$~30\fm0 in the optical, which is not feasible with
current instrumentation.
The BB hot spot spectral 
component derived from the X-ray data with 
the BB+PL model (dotted line in the Fig.~\ref{fig:spectrum}) 
all the more cannot be currently reached neither in the optical nor in UV. 
The light-grey region in Fig.~\ref{fig:spectrum} contains all possibilities 
for the soft thermal component in the BB+BB+PL model for allowed $R^{\infty}$ range
in accordance with the 99\% confidence contour 
of Fig.~\ref{fig:weiss}. 
We may conclude that the optical upper limit
does not put any additional constraints on thermal emission from the NS surface.

However, according to  Fig.~\ref{fig:spectrum}, 
the entire surface thermal spectral component  
can be reached in UV. It can also dominate over the PL component there, if 
the PL component has approximately flat spectral slope from the optical to the UV, 
as it is observed for other middle-aged pulsars. The latter
would be better constrained at longer optical wavelengths, 
less affected by the interstellar extinction.
%The contribution from the  hot spot component, if any, is negligible in the UV. 
Therefore, UV observations of J0357$+$3205
would be useful to constrain its surface temperature. 
There are only few pulsars with thermal emission detected in the UV range
namely PSR B0656+14 
\citep{Durant}, PSR B1055$-$52 \citep{Mignani1055}, PSR J0437$-$4715 \citep{Karg2004}, and 
Geminga \citep{kargaltsev2005ApJ}. In addition, \citet{Kaplan} reported 
detection of UV thermal emission from 
a few isolated neutron stars.    
In all these cases the UV data on thermal emission were 
of a great complement to the X-ray data.

Accounting for the direction of the pulsar proper motion and the spindown age, 
we find its likely birth place in the $\lambda$-Orionis cluster, a 5 Myr active star
forming region located in $\sim$~32\degs\ from the pulsar and
in $\sim$~450$\pm$50 pc from Earth \citep{mayne2008MNRAS}.
Several authors proposed that an expanding molecular ring surrounding the 
cluster is a supernova remnant left by a Type
II supernova explosion of a massive companion of the
O-type $\lambda$~Ori star about 1 Myr ago \citep[see e.g.,][]{cunha1996AsAp,dolan2002AJ}. 
Adopting this birth place we independently constrained 
the pulsar age of 0.2--1.3 Myr, accounting for 
the pulsar proper motion uncertainties, 
and the cluster and the pulsar distance ranges. 
This is consistent with its spindown age of 0.54 Myr.

%%%%%%%%%%%%%%%%%%%%%%%%%%%%%%%%%%%%%%%%%%%%%%%%%%%%%%%%%%%%%%%%%%%%%%%%%%%%%% fig. coolcurves
\label{sec4}
\begin{figure}[t]
  \begin{center}
   \includegraphics[scale=0.61]{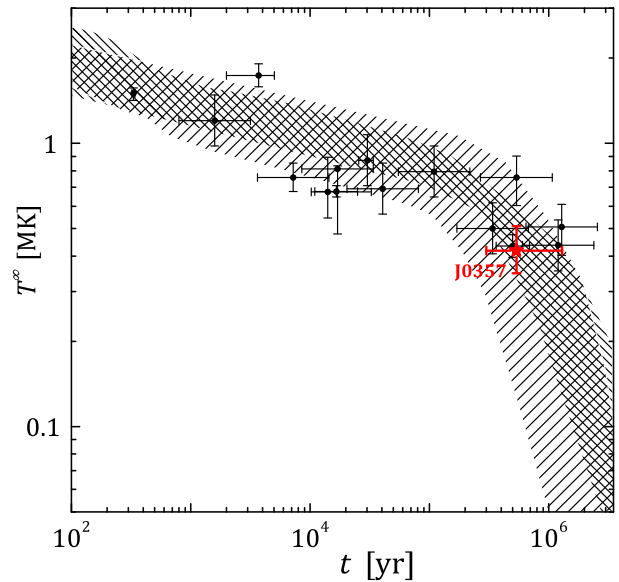}
  \end{center}
  \caption{
  Measured entire surface temperatures $T^{\infty}$ as 
  seen by a distant observer 
  (filled circles) 
  for NSs of different age $t$ in comparison with the cooling
  theory predictions (hatched regions). Dense hatched region 
  correspond to the standard cooling theory,
  while sparse hatched region shows the minimal cooling theory predictions.
  The bold star with error-bars show the 
  PSR~J0357$+$3205 surface temperature.
  }
\label{fig:cool}
\end{figure}
%%%%%%%%%%%%%%%%%%%%%%%%%%%%%%%%%%%%%%%%%%%%%%%%%%%%%%%%%%%%%%%%%%%%%%%%%%%%%%%%%%%%%%%%%%%%%%

Accepting this age range 
and the NS effective temperature of 
36$^{+8}_{-6}$ eV derived from the NSMAX+PL fit
we can compare these with the NS cooling theory predictions.
The J0357$+$3205 position on the temperature--age plane is shown in Fig.~\ref{fig:cool} 
with the bold star with error-bars. The 
data for other isolated neutron stars (filled circles) are taken from \citet{shternin2011MNRAS}. 
It is seen, that J0357$+$3205 is among the coldest cooling NSs
known. 
The dense hatched region shows the range of the NS
temperatures that can be obtained by the standard
cooling theory where the modified Urca processes 
are considered as the main neutrino
emission mechanism \citep[e.g.,][]{yakovlev2004ARA}. 
The J0357$+$3205 position agrees well with the
standard cooling theory.
However, as seen from Fig.~\ref{fig:cool} the standard cooling theory 
is insufficient to reproduce the data on all 
cooling NSs. Therefore, with sparse hatched region we show 
the range of cooling curves obtained 
within the minimal cooling scenario
\citep{gusakov2004AsAp,page2004ApJS,page2009ApJ} which takes into account the
presence of the baryon superfluidity inside neutron stars. In this scenario 
the specific process of the neutrino emission due to a
Cooper pair formation cools the star more effectively than the
modified Urca process. To date,
the parameters of the superfluidity can be plausibly adjusted  (sparse hatched region) to
fit all the data on the observed NSs temperatures \citep{gusakov2004AsAp},
including the likely rapidly cooling NS in Cas A
\citep{shternin2011MNRAS}. 
Obviously, J0357$+$3205 agrees with the minimal cooling scenario as well.  

At the same time, according to Fig.~\ref{fig:weiss}, 
the entire surface temperature in the blackbody spectral model is poorly constrained, 
taking into account the uncertainties 
in the NS radius and distance to the pulsar. Thus it is not possible 
to extract any valuable information from comparison of the BB+BB+PL fit results with the 
cooling theories.

To summarise, our deep optical observations of PSR
J0357$+$3205 allowed us to constrain  the pulsar
nonthermal emission, suggesting a strong spectral break in this emission 
between the optical and X-rays.
Reanalysis of X-ray data allowed us
to constrain the NS thermal spectrum and to measure 
the effective temperature of the NS surface $T^{\infty}$~= 36$^{+8}_{-6}$ eV.
Comparing the optical upper limit with the NS thermal spectrum
we conclude that the thermal emission 
from the entire surface of the NS can be feasibly examined in the UV
range and likely dominate there over the 
nonthermal emission and the emission from pulsar hot spot(s). 
This makes J0357$+$3205 a promising target for UV observations.

\begin{acknowledgements} 
We are grateful to anonymous referee for useful comments allowing us to improve the paper 
and to Alexander Potekhin and Dmitriy Barsukov for helpful discussion.
We also thank Antonio Cabrera Lavers for the discussion on the GTC data reduction.
The work was partially supported by CONACYT 151858 projects, 
the Russian Foundation for Basic 
Research (grants 13-02-12017-ofi-m, 14-02-00868-a), and
RF Presidential Programme MK-2837.2014.2.
\end{acknowledgements}

%%%%%%%%%%%%%%%%%%% REFERENCES %%%%%%%%%%%%%%%%%%%%%%%%%%%%%%%%%%%%%%%%%%%%%%%%%%%%%%%%
\bibliographystyle{aa}
\bibliography{ref0357}
%%%%%%%%%%%%%%%%%%%%%%%%%%%%%%%%%%%%%%%%%%%%%%%%%%%%%%%%%%%%%%%%%%%%%%%%%%%%%%%%%%%%%%%%

\end{document}